\documentclass{jaa}
\usepackage{natbib}
\usepackage{multicol}
\usepackage{graphicx}
\usepackage{multirow}
\usepackage{subfigure}
\usepackage{xcolor}
\usepackage{adjustbox}
\usepackage[colorlinks=true,linkcolor=blue,citecolor=blue]{hyperref}
\begin{document}
%\sloppy

%%paper title
%%For line breaks \\ can be used within title
\title{Detection of monothioformic acid towards the solar-type protostar \\IRAS 16293--2422}

\author{Arijit Manna\textsuperscript{1}, Sabyasachi Pal\textsuperscript{1,*}}
\affilOne{\textsuperscript{1}Department of Physics and Astronomy, Midnapore City College, Paschim Medinipur, West Bengal, India 721129\\}

%%escape two column mode for title, affiliation and abstract
%%by giving \twocolumn command as shown
\twocolumn[{
\maketitle
%%include \corres to print the corresponding author Email id
\corres{sabya.pal@gmail.com}

%%include \msinfo for
%%manuscript information such as
%%received, revised and accepted dates
%%
%\msinfo{1 January 2015}{1 January 2015}
%%abstract

\vspace{0.5cm}
\begin{abstract}
In the interstellar medium (ISM), the complex organic molecules that contain the thiol group (--SH) play an important role in the polymerization of amino acids. We look for SH-bearing molecules in the chemically rich solar-type protostar IRAS 16293--2422. After the extensive spectral analysis using the local thermodynamic equilibrium (LTE) model, we have detected the rotational emission lines of trans-isomer monothioformic acid (t-HC(O)SH) towards the IRAS 16293 B using the Atacama Large Millimeter/Submillimeter Array (ALMA). We did not observe any evidence of cis-isomer monothioformic acid (c-HC(O)SH) towards the IRAS 16293 B. The column density of t-HC(O)SH towards the IRAS 16293 B was (1.02$\pm$0.6)$\times$10$^{15}$ cm$^{-2}$ with an excitation temperature of 125$\pm$15 K. The fractional abundance of t-HC(O)SH with respect to H$_{2}$ towards the IRAS 16293 B is 8.50$\times$10$^{-11}$. The column density ratio of t-HC(O)SH/CH$_{3}$SH towards the IRAS 16293 B is 0.185. We compare our estimated abundance of t-HC(O)SH towards the IRAS 16293 B with the abundance of t-HC(O)SH towards the galactic center quiescent cloud G+0.693--0.027 and hot molecular core G31.41+0.31. After the comparison, we found that the abundance of t-HC(O)SH towards the IRAS 16293 B is several times of magnitude lower than G+0.693--0.027 and G31.41+0.31. We also discuss the possible formation mechanism of t-HC(O)SH in the ISM.
\end{abstract}

%%insert keywords separated by 3 hyphens using \keywords{words}
\keywords{ISM: individual objects (IRAS 16293--2422) -- ISM: abundances -- ISM: kinematics and dynamics -- stars: formation -- astrochemistry}

%%close the twocolumn escape here
%%include \doinum{number}for the DOI number in the header
%%include \volnum{number} for the volume number in the header
%%include \year{yyyy} for  year of publication in the header
%%include \pgrange{num--num} page range of article in the header
%%include \artcitid{num} for the article citation id
%%include \lp to print last page of the article
%%include \setcounter{page}{pagenum} for the exact starting page of the article
}]
\doinum{xyz/123}
\artcitid{\#\#\#\#}
\volnum{000}
\year{2021}
\pgrange{1--11}
\setcounter{page}{1}
\lp{11}

\section{Introduction}
\label{sec:intro} 
In the interstellar medium (ISM) and circumstellar shells, about 270 molecules have been detected at millimeter and sub-millimeter wavelengths\footnote{\url{https://cdms.astro.uni-koeln.de/classic/molecules}}. Sulfur (S) is the tenth most abundant compound in space, and so far a very small number of sulfur-bearing organic molecules have been detected in the ISM. A few astronomical objects in the ISM have been found that contain thiol (SH)-bearing molecules with more than four atoms. For example, methyl mercaptan ({CH$_{3}$SH}) has been found in a variety of astronomical environments, like the high-mass star formation region Sgr B2 \citep{lin79}, prestellar cores \citep{gib20}, low-mass protostar IRAS 16293--2422 \citep{maj16, dro18}, hot molecular core G31.41+0.31 \citep{gor21}, and galactic center quiescent cloud G+0.693--0.027 \citep{rod21}. Another SH-bearing molecule, ethyl mercaptan ({C$_{2}$H$_{5}$SH}), was tentatively detected towards the hot molecular core Orion KL and confirmed to be detected towards the G+0.693--0.027 \citep{kol14, rod21}. Earlier, other S-bearing complex organic molecules such as thioacetaldehyde ({CH$_{3}$CHS}), thioformamide ({NH$_{2}$CHS}), and S-methyl thioformate ({CH$_{3}$SC(O)H}) were not found in the star-forming regions and cold molecular clouds \citep{mar20, mot20, jab20}.

\begin{figure}
	\centering
	\includegraphics[width=0.47\textwidth]{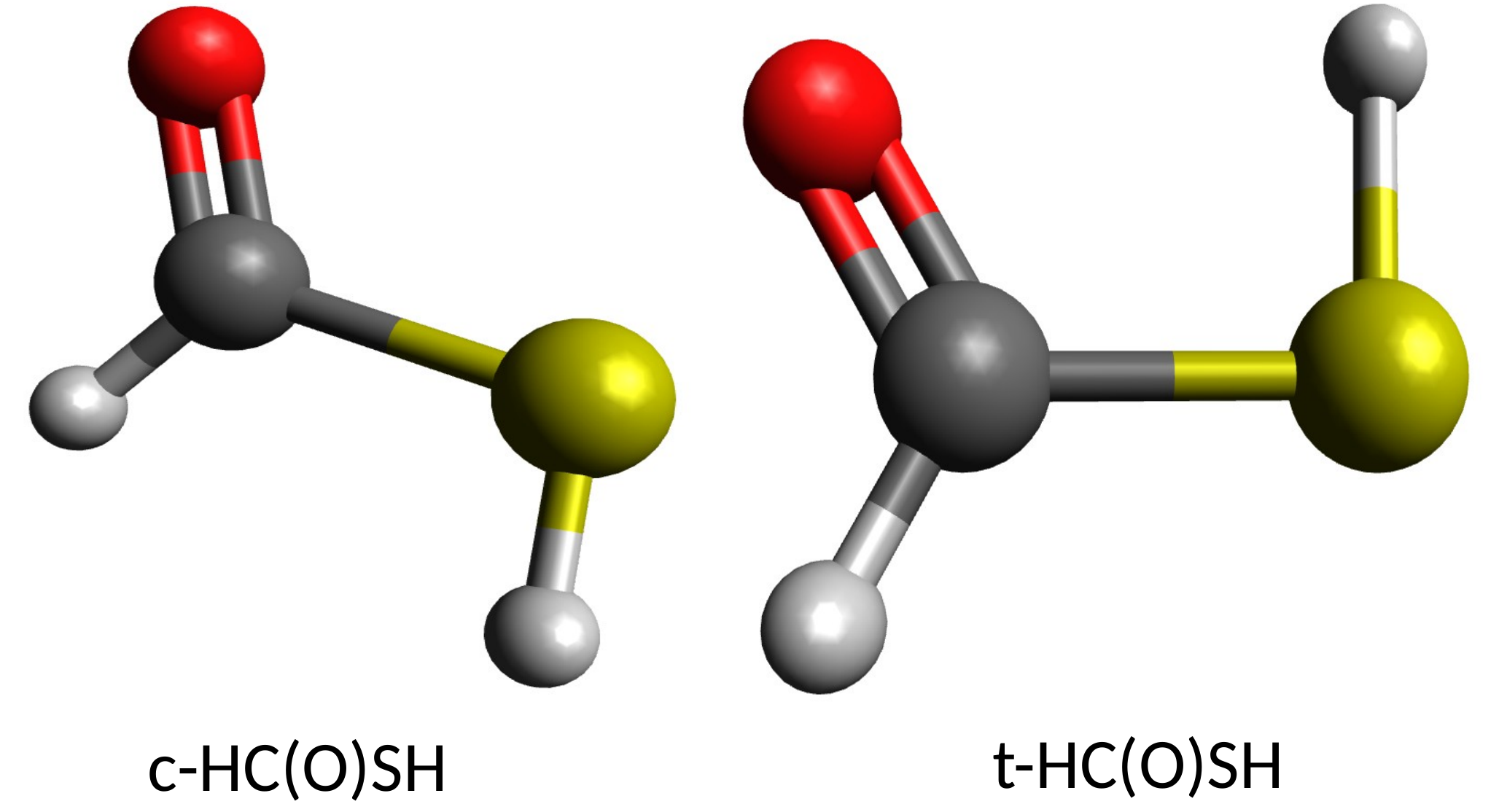}
	\caption{Three-dimensional molecular structure of c-HC(O)SH and t-HC(O)SH. The grey atoms are carbon (C), the red atoms are oxygen (O), the white atoms are hydrogen (H), and the yellow atoms are sulfur (S) \citep{gar22}.}
	\label{fig:molecule}
\end{figure}

Monothioformic acid (HC(O)SH) is known as one of the rare SH-bearing molecule that was first detected towards the G+0.693-0.027 \citep{rod21}. The HC(O)SH molecule has two isomers, i.e., the cis-isomer (c-HC(O)SH) and the trans-isomer (t-HC(O)SH). The 3D molecular structure of c-HC(O)SH and t-HC(O)SH is shown in Figure~\ref{fig:molecule}. The relative electronic energy between c-HC(O)SH and t-HC(O)SH is 0.68 kcal mol$^{-1}$ \citep{gar22}. The dipole moments of the t-HC(O)SH are $\mu_{a}$ = 1.366 D (a-type) and $\mu_{b}$ = 0.702 D (b-type) \citep{hok76}. Similarly, the dipole moments of c-HC(O)SH are $\mu_{a}$ = 1.805 D (a-type) and $\mu_{b}$ = 2.228 D (b-type) \citep{hok76}.  \citet{rod21} detected the rotational emission lines of t-HC(O)SH from the G+0.693--0.027, but the author could not observe any evidence of c-HC(O)SH. Recently, the rotational emission lines of t-HC(O)SH were detected towards the chemically rich hot molecular core G31.41+0.31 using the ALMA with an estimated abundance of 1.4$\times$10$^{-8}$ but they could not identify any non-blended emission lines of c-HC(O)SH towards the G31.41+0.31 \citep{gar22}.

IRAS 16293--2422 (hereafter IRAS 16293) is classified as a class 0 protostellar core, which is located in the L1689 region of the $\rho$ Ophiuchus (hereafter $\rho$ Oph) cloud \citep{tac00, nut06, you06, pad08}. The distance of $\rho$ Oph cloud is $\sim$120 pc \citep{lom08}. The luminosity of IRAS 16293 is 21$\pm$5 \(\textup{L}_\odot\) \citep{cor04}. IRAS 16293 is known as a binary object, which consists of two solar-like protostars, IRAS 16293 A and IRAS 16293 B. The protostars IRAS 16293 A and IRAS 16293 B are surrounded by chemically rich hot corinos \citep{caz03, bot04}. Two protostars, IRAS 16293 A and IRAS 16293 B, are separated by $\sim$5$^{\prime\prime}$ ($\sim$600 AU) from each other, and their masses are $\sim$0.5 \(\textup{M}_\odot\) \citep{wot89, mun92, bot04, lo00}. IRAS 16293 B contains more complex organic molecules compared to IRAS 16293 A \citep{bot04, kun04, pin12}. The molecular line width towards the IRAS 16293 B is much narrower (1--1.5 km s$^{-1}$) than the IRAS 16293 A ($\geq$3.0 km s$^{-1}$) because a disk/envelope system is observed in the IRAS 16293 B, which has nearly face-on geometry in contrast to the edge-on geometry of the IRAS 16293 A \citep{pin12, zap13, oy16}. Additionally, an inverse P-Cygni profile is observed towards the IRAS 16293 B, which indicates the existence of the infalling gas in front of the protostar along the line of sight \citep{pin12}. Previously, many complex organic molecules have been identified towards the IRAS 16293 i.e., cyanamide (NH$_{2}$CN) \citep{cou18}, formamide ({NH$_{2}$CHO}) \citep{kan16, cou16}, methyl isocyanate ({CH$_{3}$NCO}) \citep{lig17}, ethylene glycol (({CH$_{2}$OH})$_{2}$) \citep{jo12}, and the sugar-like molecule glycolaldehyde ({CH$_{2}$OHCHO}) \citep{jo12}, etc.

In this article, we present the first detection of the rotational emission lines of the SH-bearing molecule t-HC(O)SH towards the IRAS 16293 B using the ALMA. The ALMA observations and data reduction are presented in Section~\ref{obs} The result of the identification of the rotational emission lines of t-HC(O)SH is shown in Section~\ref{res} The discussion and conclusion of the detection of t-HC(O)SH towards the IRAS 16293 B are shown in Section~\ref{dis} and \ref{conclu}

\section{Observations and data reductions}
\label{obs}

The solar-type protostar IRAS 16293 was observed to study the rotational emission lines of $^{16}$O$^{18}$O using the ALMA band 6 with the 12 m array antennas of cycle 4 (project id: 2016.1.01150.S, PI: Taquet, Vianney). The observation was done on November 10, 2016, with an on-source integration time of 2.654 hour. The observed phase center of IRAS 16293 was ($\alpha,\delta$)$_{\rm J2000}$ = 16:32:22.720, --24:28:34.300. During the observation, the flux and the bandpass calibrator were taken as J1527--2422 and the phase calibrator was taken as J1625--2527. A total of forty antennas were used to observe the IRAS 16293 with a minimum baseline of 15.1 m and a maximum baseline of 1062.5 m. The observation was taken using four spectral windows with frequency ranges of 233.712--234.180 GHz, 234.918--235.385 GHz, 235.908--236.379 GHz, and 236.368--236.841 GHz with a spectral resolution of 244 kHz and bandwidth 469 MHz.

\begin{figure*}
	\centering
	\includegraphics[width=0.92\textwidth]{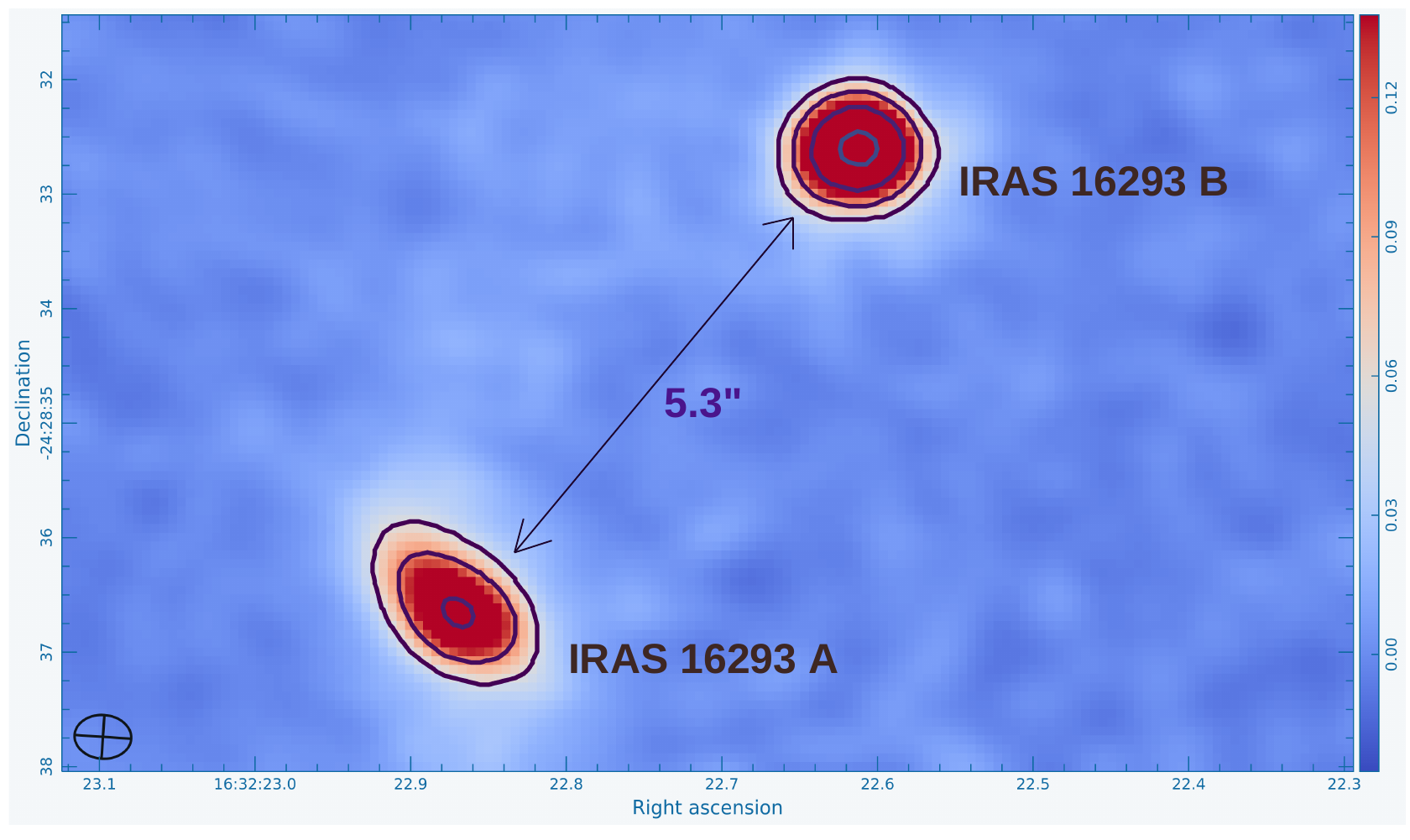}
	\caption{Millimeter-wavelength (1.274 mm) continuum emission image of IRAS 16293--2422, a solar-like protostar with two sources, IRAS 16293 A and IRAS 16293 B. The black circle is the synthesised beam of the continuum emission image. The synthesised beam size of the continuum image is 0.498$^{\prime\prime}\times$ 0.381$^{\prime\prime}$. The blue contour levels are started at 3$\sigma$, where $\sigma$ is the RMS of the continuum emission image, which is 3.85 mJy.}
	\label{fig:continuum}
\end{figure*}

We have used the Common Astronomy Software Application ({\tt CASA 5.4.1}) for the data reduction using the ALMA data reduction pipeline \citep{mc07}.  For flux calibration, we have used the Perley-Butler 2017 flux calibration model for each baseline using the task {\tt SETJY} \citep{pal17}. To construct the flux and bandpass calibration after the flagging of the bad antenna data, we applied the pipeline tasks {\tt hifa\_bandpassflag} and {\tt hifa\_flagdata}. After the initial data reduction, we used the CASA task {\tt MSTRANSFORM} with all available rest frequencies to separate the target source IRAS 16293. We constructed the continuum images of the IRAS 16293 from the line-free channels using the CASA task {\tt TCLEAN}. For the continuum subtraction operation, we have used the task {\tt UVCONTSUB} in the UV plane of separated calibrated data. We used the task {\tt TCLEAN} with a Briggs weighting robust value of 0.5 to create the spectral images of IRAS 16293. During the production of the spectral line images, we used the {\tt SPECMODE=CUBE} parameter in the task {\tt TCLEAN}. Finally, we have used the task {\tt IMPBCOR} for the correction of the primary beam pattern in the continuum and spectral images.

\section{Result}
\label{res}
\subsection{Continuum emission towards the IRAS 16293}
 
We present the millimeter-wavelength continuum emission image of IRAS 16293 at frequency 235.29 GHz (1.274 mm) in Figure~\ref{fig:continuum}. After the generation of the continuum emission image of IRAS 16293 using the CASA task {\tt TCLEAN} with the hogbom deconvolver, we fitted a 2D Gaussian using the task {\tt IMFIT} over the continuum emission image of sources IRAS 16293 A and IRAS 16293 B to estimate the physical parameters of the image. For source IRAS 16293 A, the estimated integrated flux density is 743.6$\pm$23 mJy and the peak flux density is 207.1$\pm$5 mJy beam$^{-1}$. Similarly, for source IRAS 16293 B, the integrated flux density is 934.5$\pm$22 mJy and the peak flux density is 464.1$\pm$7.9 mJy beam$^{-1}$. We noticed that the continuum emission regions of sources IRAS 16293 A and IRAS 16293 B are larger than the synthesised beam size of 0.498$^{\prime\prime}\times$ 0.381$^{\prime\prime}$. That indicates IRAS 16293 A and IRAS 16293 B are resolved at wavelength 1.274 mm.

\begin{figure*}
	\centering
	\includegraphics[width=1.0\textwidth]{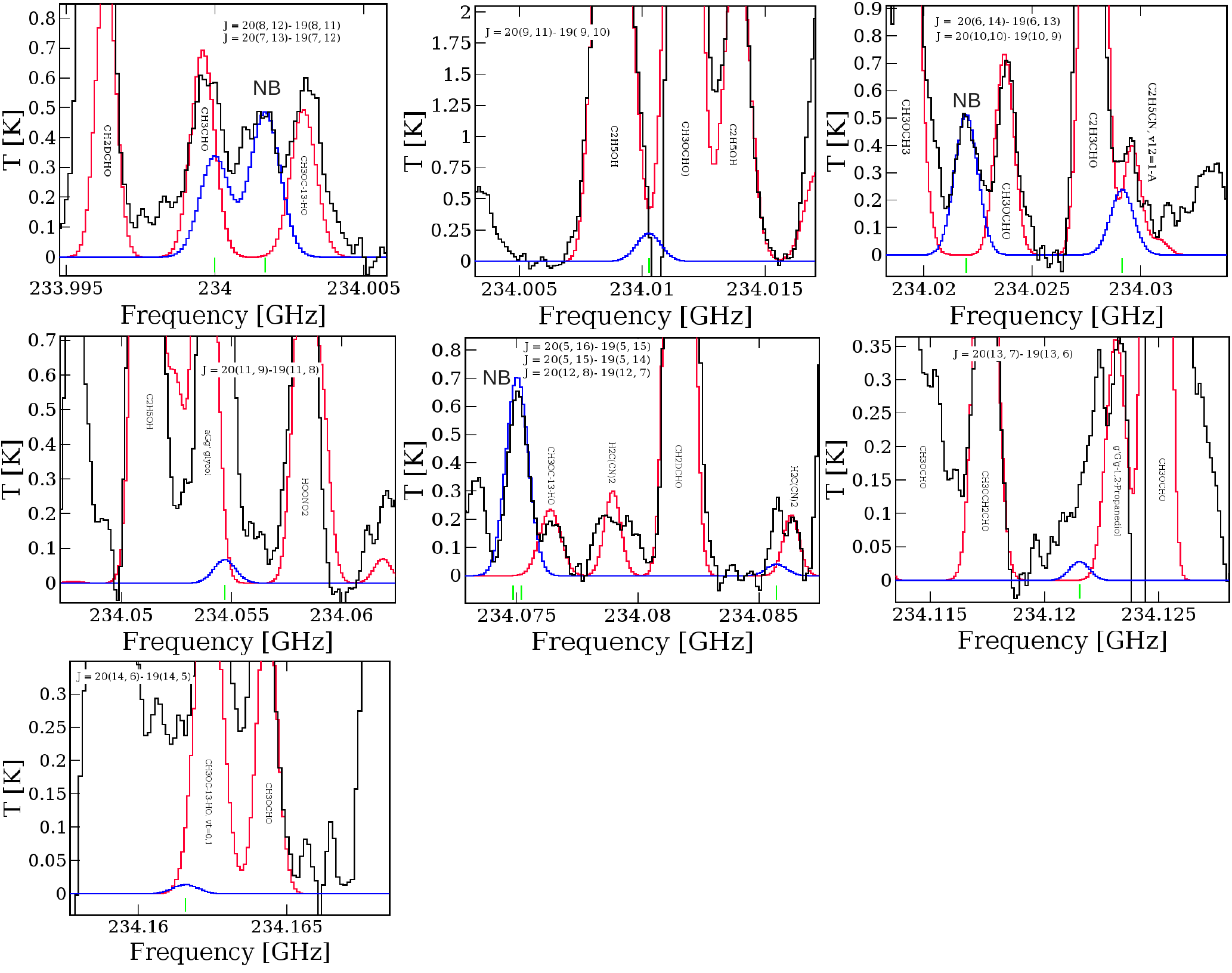}
	\caption{Detected rotational emission lines of t-HC(O)SH towards the IRAS 16293 B with different transitions. The black spectrum presents the observed millimeter-wavelength molecular spectra of IRAS 16293 B. The blue synthetic spectra present the LTE model of t-HC(O)SH, and the red spectra indicate the LTE model of other nearby molecular transitions. The green vertical lines indicate the rest frequency positions of the detected transitions of t-HC(O)SH. In the emission spectra, `NB' indicate the non-blended emission lines of t-HC(O)SH.}
	\label{fig:emissionspectra}
\end{figure*}

\begin{table*}
	%\begin{minipage}[t]{\columnwidth}
	\centering
	%\scriptsize 
	\caption{Summary of the LTE fitted line parameters of the t-HC(O)SH towards the IRAS 16293 B.}
	\begin{adjustbox}{width=1.0\textwidth}
		\begin{tabular}{ccccccccccccccccc}
			\hline 
			Observed frequency &Transition & $E_{u}$ & $A_{ij}$ &g$_{up}$&FWHM &Optical depth& V$_{LSR}$ &Remark\\
			
			(GHz) &(${\rm J^{'}_{K_a^{'}K_c^{'}}}$--${\rm J^{''}_{K_a^{''}K_c^{''}}}$) &(K)&(s$^{-1}$) & &(km s$^{-1}$) &($\tau$)&(km s$^{-1}$) & \\
			\hline
			233.999$^{*}$&20(8,12)--19(8,11)&290.26&1.14$\times$10$^{-4}$&41&1.352$\pm$0.14&1.302$\times$10$^{-3}$&2.760&Blended with CH$_{3}$CHO\\
			
			234.001$^{*}$&20(7,13)--19(7,12)&249.91&1.19$\times$10$^{-4}$&41&1.353$\pm$0.30&1.556$\times$10$^{-3}$&2.761&Non blended\\
			
			234.010$^{*}$&20(9,11)--19(9,10)&335.97&1.08$\times$10$^{-4}$&41&1.349$\pm$0.40&1.062$\times$10$^{-3}$&2.723&Blended with C$_{2}$H$_{5}$OH\\
			
			234.021$^{*}$&20(6,14)--19(6,13)&214.92&1.24$\times$10$^{-4}$&41&1.352$\pm$0.70&1.813$\times$10$^{-3}$&2.710&Non blended\\
			
			234.029$^{*}$&20(10,10)--19(10,9)&387.01&1.02$\times$10$^{-4}$&41&1.351$\pm$0.35&8.422$\times$10$^{-4}$&2.732&Blended with C$_{2}$H$_{5}$CN\\
			
			234.054$^{*}$&20(11,9)--19(11,8)&443.38&9.48$\times$10$^{-5}$&41&1.359$\pm$0.21&6.491$\times$10$^{-4}$&2.705&Below 2$\sigma$ and Blended with  aGg$^{\prime}$glycol\\
			
			234.074&20(5,16)--19(5,15)&185.30&1.27$\times$10$^{-4}$&41&1.342$\pm$0.12&2.062$\times$10$^{-3}$&2.725&Non blended\\
			
			234.075&20(5,15)--19(5,14)&185.30&1.27$\times$10$^{-4}$&41&1.353$\pm$0.15&2.071$\times$10$^{-3}$&2.725&Non blended\\
			
			234.085$^{*}$&20(12,8)--19(12,7)&505.06&8.70$\times$10$^{-5}$&41&1.352$\pm$0.78&3.506$\times$10$^{-3}$&2.705&Below noise level and blended\\
			
			234.121$^{*}$&20(13,7)--19(13,6)&572.03&7.85$\times$10$^{-5}$&41&1.356$\pm$0.98&3.502$\times$10$^{-4}$&2.715&Below noise level and blended\\
			
			234.161$^{*}$&20(14,6)--19(14,5)&644.28&6.94$\times$10$^{-5}$&41&1.351$\pm$0.98&2.403$\times$10$^{-4}$&2.709&Blelow noise level and blended\\
			
			\hline
		\end{tabular}	
	\end{adjustbox}
	\label{tab:MOLECULAR DATA}\\
	{\color{blue}{*}}--The transition of t-HC(O)SH contain double with frequency difference  $\leq$100 kHz. The second transition is not shown.\\
	%		\end{minipage}[t]{\columnwidth}
\end{table*}

\subsection{Identification of emission lines of t-HC(O)SH towards the IRAS 16293 B}
 
We create the millimeter-wavelength molecular emission spectra from the continuum-subtracted spectral images of IRAS 16293 by using a 2.07$^{\prime\prime}$ diameter circular region over the hot corino protostar object IRAS 16293 B. The synthesised beam size of all spectral images is 0.546$^{\prime\prime}\times$0.391$^{\prime\prime}$. The phase center of IRAS 16293 B is RA (J2000) = 16$^{h}$32$^{m}$22$^{s}$.603, Dec (J2000) = --24$^\circ$28$^{\prime}$32$^{\prime\prime}$.729. We extract the millimeter-wavelength molecular spectra from IRAS 16293 B because IRAS 16293 B is highly chemically rich compared to IRAS 16293 A. The spectral line width of IRAS 16293 B is $\sim$1.0--1.5 km s$^{-1}$ but the spectral line width of IRAS 16293 A is $\geq$3.0 km s$^{-1}$ \citep{jo16}. So, the searching of complex molecular species from the broad spectral lines towards the IRAS 16293 A is difficult, and there are high chances of detecting the blended spectral lines rather than the non-blended spectral lines \citep{jo12, jo16}. The systematic velocity ($V_{LSR}$) of the IRAS 16293 B is 2.7 km s$^{-1}$ \citep{jo16, taq18}.

After the extraction of the molecular spectral lines of IRAS 16293 B, we fit the first-order polynomial over the molecular spectra for baseline correction using the GNU R package {\tt hyperSpec}\footnote{\url{https://cran.r-project.org/web/packages/hyperSpec/vignettes/baseline.pdf}}. We used the local thermodynamic equilibrium (LTE) model with the Cologne Database for Molecular Spectroscopy (CDMS) to identify the molecular emission lines of t-HC(O)SH (CDMS entry 062515) towards the IRAS 16293 B \citep{mu05}. For the LTE modelling, we have used the CASSIS\footnote{CASSIS has been developed by IRAP-UPS/CNRS} \citep{vas15}. The LTE assumptions are reasonable in the inner region of IRAS 16293 B because the gas density of the warm inner region of the low-mass protostellar envelopes is $\geq$10$^{10}$ cm$^{-2}$ \citep{jo16, cou18}. For fitting the LTE model over the molecular spectra of IRAS 16293 B, we used the Markov Chain Monte Carlo (MCMC) algorithm in CASSIS. For running the MCMC algorithm, we used the Python script interface in CASSIS. During the LTE modelling in the molecular spectra of IRAS 16293 B, we consider T$_{c}$ = 0 and the cosmic background temperature of T$_{bg}$ = 2.73 K\footnote{Similar method was also used by \citet{taq18}.}. The LTE-fitted rotational emission spectra of t-HC(O)SH are shown in Figure~\ref{fig:emissionspectra}. After the LTE modelling, we detected eleven transition lines of t-HC(O)SH within the observable frequency ranges. The upper-level energies ($E_{u}$) of the detected eleven transitions of t-HC(O)SH vary from 185.30 K to 644.28 K. Among the detected eleven transition lines of t-HC(O)SH, only four transition lines are non-blended, and those non-blended lines are identified as higher than 5$\sigma$. The upper-level energies ($E_{u}$) of the non-blended transitions of t-HC(O)SH vary from 185.30 K to 249.91 K. After the LTE analysis, we observed that the J = 20(5,16)--19(5,15) and J = 20(5,15)--19(5,14) transition lines of t-HC(O)SH at frequencies of 234.074 GHz and 234.075 GHz are present in the single spectral profile which means those two transitions of t-HC(O)SH are blended with each other. We also found that the LTE spectra of J = 20(12,8)--19(12,7), J = 20(13,7)--19(13,6), and J = 20(14,6)--19(14,5) transition lines of t-HC(O)SH at frequencies 234.085 GHz, 234.121 GHz, and 234.161 GHz did not generate properly (below 1.5 $\sigma$) due to high upper state energies ($E_{u}$) of 505.06 K, 572.03 K, and 644.28 K. There were no missing any high-intensity transitions of t-HC(O)SH between the observable frequency ranges. The LTE-fitted spectral line parameters of t-HC(O)SH are shown in Table~\ref{tab:MOLECULAR DATA}. After the LTE analysis, we noticed that all observed non-blended emission lines of t-HC(O)SH are properly fitted with modelled spectra, but the blended lines are not properly fitted. At the full-beam offset position, the best-fit column density of t-HC(O)SH is (1.02$\pm$0.6)$\times$10$^{15}$ cm$^{-2}$ with an excitation temperature of 125$\pm$15 K and source size of 0.5$^{\prime\prime}$. Our derived excitation temperature indicates that the detected transition lines of t-HC(O)SH arise from the warm inner region of IRAS 16293 B because the temperature of the hot corinos is above 100 K \citep{jo12, dro18, cou18}. The full-width half maximum (FWHM) of the LTE model spectra of t-HC(O)SH is 1.35$\pm$0.3 km s$^{-1}$. The derived excitation temperature of t-HC(O)SH agrees with the reported excitation temperature of another SH-bearing molecule, CH$_{3}$SH, by \citet{dro18}. Additionally, we have also searched more than 200 different molecules in that data set, including all detected molecules towards the IRAS 16293 B by \cite{taq18}. We applied the LTE modelling of other simple and complex molecules towards the IRAS 16293 B to understand the line contamination by other molecules with t-HC(O)SH. The produced total LTE spectra of different molecules are shown with red lines in Figure~\ref{fig:emissionspectra}. After the detection of t-HC(O)SH, we also looked for the rotational emission lines of c-HC(O)SH towards the IRAS 16293 B, but no clear transition of t-HC(O)SH is present in the data.

\subsection{Fractional abundance of t-HC(O)SH towards the IRAS 16293 B}

The fractional abundance of t-HC(O)SH with respect to H$_{2}$ towards the IRAS 16293 B is 8.50$\times$10$^{-11}$, while the column density of molecular H$_{2}$ towards the IRAS 16293 B is 1.20$\times$10$^{25}$ cm$^{-2}$ \citep{jo16, cou18}. The fractional abundance of t-HC(O)SH towards the IRAS 16293 B with respect to another SH--bearing molecule, CH$_{3}$SH is 0.185, where the column density of CH$_{3}$SH towards the IRAS 16293 B is 5.5$\times$10$^{15}$ cm$^{-2}$ \citep{dro18}. We estimate the molecular column density ratio between t-HC(O)SH and CH$_{3}$SH because both molecules share --SH as a common precursor. Our derived fractional abundance of t-HC(O)SH (8.50$\times$10$^{-11}$) is lower than the abundance of CH$_{3}$SH (4.58$\times$10$^{-10}$)\footnote{The abundance of CH$_{3}$SH towards the IRAS 16293 B was reported by \citet{dro18}}. This result indicates that the abundance roughly follows the monotonic decrease from CH$_{3}$SH to t-HC(O)SH.

We also compared the estimated abundance of t-HC(O)SH towards the IRAS 16293 B with the G+0.693--0.027 and G31.41+0.31. Earlier, \citet{rod21} and \citet{gar22} detected the rotational emission lines of t-HC(O)SH from G+0.693--0.027 and G31.41+0.31, with an estimated fractional abundance 1.2$\times$10$^{-10}$ and 1.4$\times$10$^{-8}$ respectively. Our estimated abundance of t-HC(O)SH towards the IRAS 16293 B is 8.50$\times$10$^{-11}$, which is several times of magnitude lower than the abundance of t-HC(O)SH towards the G+0.693–0.027 and G31.41+0.31. That comparison also indicates that the chemical richness of SH-bearing molecules towards the IRAS 16293 B is less than G+0.693--0.027 and G31.41+0.31. 

\begin{table}
	%\begin{minipage}[t]{\columnwidth}
	\centering
	%\scriptsize 
	\caption{The upper limit column density of the other SH-bearing molecules towards IRAS 16293 B.}
	\begin{adjustbox}{width=0.48\textwidth}
		\begin{tabular}{ccccccccccccccccc}
			\hline 
			Molecule&CDMS entry & Upper limit column density\\
			&           & (cm$^{-2}$)\\
			\hline
			~~a-C$_{2}$H$_{3}$SH& 060522 &$\leq$(2.32$\pm$0.2)$\times$10$^{12}$\\	
			~~s-C$_{2}$H$_{3}$SH&060521&$\leq$(1.53$\pm$0.8)$\times$10$^{12}$ \\
			~~a-C$_{2}$H$_{5}$SH&062524&	$\leq$(7.82$\pm$0.6)$\times$10$^{11}$\\ 
			~~g-C$_{2}$H$_{5}$SH&062523&$\leq$(3.05$\pm$0.9)$\times$10$^{11}$\\
			c-HC(O)SH&062516&$\leq$(4.50$\pm$0.8)$\times$10$^{14}$\\
			c-HC(S)SH&078507& $\leq$(5.21$\pm$0.7)$\times$10$^{14}$ \\
			t-HC(S)SH&078506& $\leq$(3.82$\pm$0.4)$\times$10$^{13}$\\
			~~~~~~~~CaSH&073504&$\leq$(6.51$\pm$0.9)$\times$10$^{11}$\\
			~~~~~HCCSH&058520&$\leq$(3.97$\pm$0.3)$\times$10$^{12}$\\
			~~~~~~~~~~KSH&072506&$\leq$(5.91$\pm$0.7)$\times$10$^{11}$\\
			~~~~~~~~MgSH&057516& $\leq$(3.46$\pm$0.4)$\times$10$^{12}$\\
			~~~~~~~~~NaSH&056522&$\leq$(9.72$\pm$0.7)$\times$10$^{10}$\\
			~~~~~~~~~~AlSH&060520&$\leq$(2.32$\pm$0.5)$\times$10$^{11}$\\
			
			\hline
		\end{tabular}	
	\end{adjustbox}
	\label{tab:other molecule}\\
	%		\end{minipage}[t]{\columnwidth}
\end{table}

\subsection{Searching of other SH-bearing molecules towards the IRAS 16293 B}
After the identification of two SH-bearing molecules i.e., t-HC(O)SH (this paper) and CH$_{3}$SH \citep{maj16, dro18} towards IRAS 16293 B, we also looked for other SH-bearing molecules using the CDMS molecular database. Using the LTE model spectra, we try to search the rotational emission and absorption lines of a-C$_{2}$H$_{3}$SH (anti-conformer vinyl mercaptan), s-C$_{2}$H$_{3}$SH (syn-conformer vinyl mercaptan), a-C$_{2}$H$_{5}$SH (anti-conformer ethyl mercaptan or ethanethiol), g-C$_{2}$H$_{5}$SH (gauche-conformer ethyl mercaptan), C-HC(S)SH (cis-isomer dithioformic acid), t-HC(S)SH (trans-isomer dithioformic acid), CaSH (calcium monohydrosulfide), HCCSH (ethynethiol), KSH (potassium hydrosulfide), MgSH (magnesium monohydrosulfide), NaSH (sodium hydrosulfide), and AlSH (aluminium monohydrosulfide) from the spectra of the IRAS 16293 B. We did not find any evidence of those SH-bearing molecules towards the IRAS 16293 B within the limit of LTE analysis. The estimated upper limit column densities of those molecules towards the IRAS 16293 B are shown in Table~\ref{tab:other molecule}. We can not perform the search of emission lines of c-HC(O)SH (cis isomer monothioformic acid) because we can not find any transitions of c-HC(O)SH in this data set. So we estimate the upper limit column density of c-HC(O)SH with respect to c-HC(S)SH. Among the listed complex SH-bearing molecules, only C$_{2}$H$_{5}$SH was earlier detected towards Orion KL and G+0.693--0.027 \citep{kol14, rod21}.

 \begin{figure*}
	%\centering
	\includegraphics[width=0.97\textwidth]{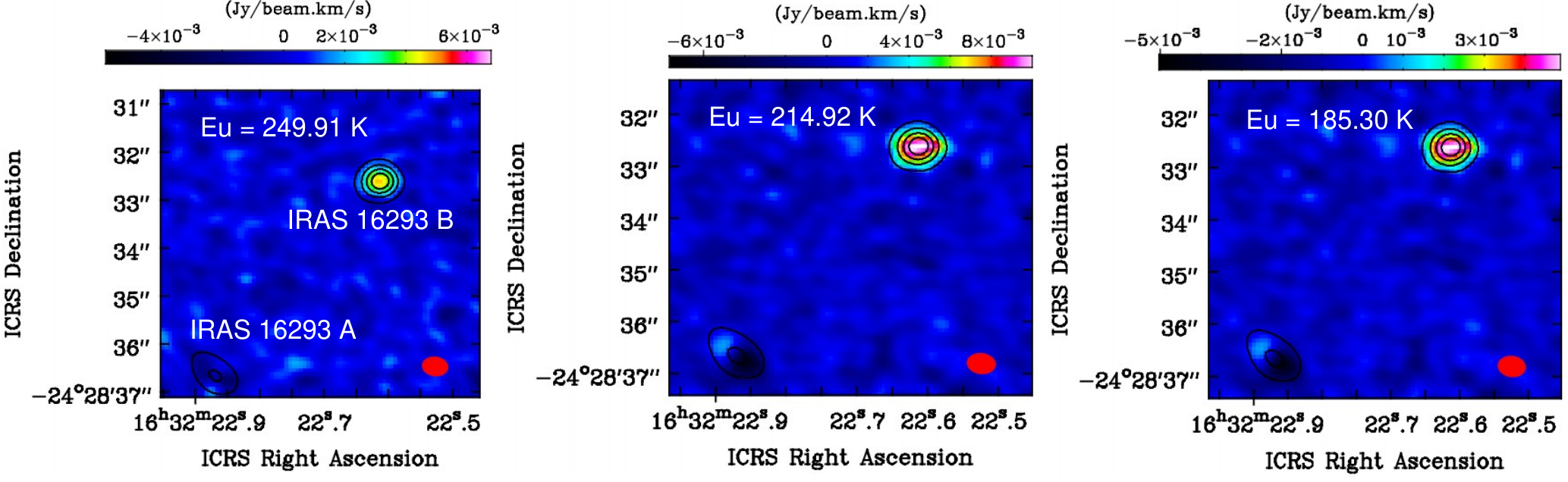}
	\caption{Integrated emission maps (moment zero maps) of t-HC(O)SH towards the IRAS 16293 B. The integrated emission maps are overlaid with the 1.274 mm continuum emission map (black contour). The contour levels are started at 3$\sigma$, where $\sigma$ = 4.23 mJy beam$^{-1}$. The red circle is the synthesised beam of the integrated emission maps. The synthesised beam size of all integrated emission maps is 0.546$^{\prime\prime}\times$0.391$^{\prime\prime}$.}
	
	\label{fig:emi}
\end{figure*}

\subsection{Spatial distribution of t-HC(O)SH }
We produce the integrated emission maps (moment zero maps) of t-HC(O)SH towards the IRAS 16293 B using the task {\tt IMMOMENTS} in CASA. During the run of the task {\tt IMMOMENTS}, we use the channel ranges of the spectral images where the emission lines of t-HC(O)SH are detected. We create the integrated emission maps of t-HC(O)SH at frequencies of 234.001 GHz, 234.021 GHz, and 234.074 GHz towards the IRAS 16293 B. The resultant integrated emission maps are shown in Figure~\ref{fig:emi}. We overlaid the 1.274 mm continuum emission map of IRAS 16293 over the integrated emission maps of t-HC(O)SH. We also found that the integrated emission maps of t-HC(O)SH exhibit a peak at the position of the continuum. The integrated emission maps indicate that the t-HC(O)SH emission lines originate from the high-density, warm inner region of IRAS 16293 B. The emission map clearly shows that the emission lines of t-HC(O)SH arise from IRAS 16293 B and not from IRAS 16293 A. To determine the emitting regions of t-HC(O)SH, we use the CASA task {\tt IMFIT} to fit the 2D Gaussian over the integrated emission maps of t-HC(O)SH. The following equation is used to calculate the t-HC(O)SH emitting regions:

\begin{equation}		
\theta_{S}=\sqrt{\theta^2_{50}-\theta^2_{\text{beam}}}		
\end{equation}
where $\theta_{\text{beam}}$ denotes the half-power width of the synthesised beam, and $\theta_{50} = 2\sqrt{A/\pi}$ denotes the diameter of the circle whose area surrounds the $50\%$ line peak of t-HC(O)SH \citep{riv17}. The derived emitting regions of the t-HC(O)SH at frequencies 234.001 GHz, 234.021 GHz, and 234.074 GHz are 0.540$^{\prime\prime}$, 0.552$^{\prime\prime}$, and 0.550$^{\prime\prime}$ respectively. We notice that the estimated emitting regions of t-HC(O)SH are comparable to or slightly greater than the synthesised beam sizes of the integrated emission maps. This demonstrates that the detected t-HC(O)SH transition lines are not spatially resolved or are only marginally resolved towards the IRAS 16293 B. As a result, determining the morphology of the spatial distribution of t-HC(O)SH towards the IRAS 16293 B is impossible. Higher spatial and angular resolution observations are required to understand the spatial distribution of t-HC(O)SH towards the IRAS 16293 B.

\section{Discussion}
\label{dis}
We first detect the emission lines of t-HC(O)SH towards the IRAS 16293 B using the ALMA. There is no information available about the formation mechanism of HC(O)SH using the gas-phase and grain-surface chemical reactions in the Kinetic Database for Astrochemistry (KIDA) \citep{wak12} and UMIST 2012 \citep{mce13} astrochemistry chemical reaction network. We assume that the astrochemical formation pathways of the SH-bearing molecules are similar to those of the OH-bearing molecules, where the oxygen atom was replaced by a sulfur atom. A possible HC(O)SH formation pathway could be replicated by HC(O)OH, in which the O atom is replaced by an S atom \citep{rod21}. Previously, \citet{io11} claimed that the HC(O)OH molecule can be formed by the reaction of CO and the radical OH in the ice. Similarly, the possible formation mechanism of HC(O)SH is,\\\\
CO + SH$\rightarrow$HSCO~~~~~~~~~~~~~~~~(1) \\\\
HSCO + H$\rightarrow$HC(O)SH~~~~~~~(2)\\\\
Reaction 1 was first proposed by \citet{ad10}. According to reaction 2, the hydrogenation of HSCO may create HC(O)SH in the ISM \citep{rod21}. Similarly, the other possible formation routes of HC(O)SH are,\\\\
HCO + SH$\rightarrow$HC(O)SH~~~~~~~(3)\\\\
and\\\\
OCS + 2H$\rightarrow$HC(O)SH~~~~~~~(4)\\\\
Reaction 3 was first proposed by \citet{rod21}, where SH can be created on the surfaces from S+H and H+H$_{2}$S via the tunnelling \citep{vi17}. Reaction 4 was proposed by \citet{rod21}, where hydrogenation of carbonyl sulfide (OCS) produced HC(O)SH in the ice. Earlier, both reactant molecules, HCO and OCS, were found towards the IRAS 16293 B \citep{riv19, dro18}. Recently, \citet{mol21} claimed that reaction 4 is the most efficient pathway to the formation of t-HC(O)SH towards the molecular cloud and high gas density hot molecular cores using the density functional theory (DFT). We recommend the astrochemistry community to use reactions 2, 3, and 4 in the three-phase warm-up model similar to \citet{gar13} and \citet{cou18} to understand the modelled abundance and proper formation routes of t-HC(O)SH towards the IRAS 16293 B. We also recommend that the astrochemistry community search for the stronger $K_{a}$ = 0,1 lines of t-HC(O)SH to identify more non-blended emission lines of t-HC(O)SH towards IRAS 16293 B.

\section{Conclusion}
\label{conclu}
In this article, we present the first detection of the SH-bearing molecule t-HC(O)SH towards the hot corino object IRAS 16293 B. The derived column density of t-HC(O)SH is (1.02$\pm$0.6)$\times$10$^{15}$ cm$^{-2}$ with an excitation temperature of 125$\pm$15 K. The fractional abundance of t-HC(O)SH with respect to H$_{2}$ towards the IRAS 16293 B is 8.50$\times$10$^{-11}$. The estimated t-HC(O)SH/{CH$_{3}$SH} ratio towards the IRAS 16293 B is 0.185. We compare our estimated abundance of t-HC(O)SH towards the IRAS 16293 B with the abundance of t-HC(O)SH towards the hot molecular core G31.41+0.31 and galactic center quiescent cloud G+0.693--0.027. After the comparison, we observed the abundance of t-HC(O)SH towards the IRAS 16293 B is several times of magnitude lower than G+0.693--0.027 and G31.41+0.31. After the identification of t-HC(O)SH (this paper) and CH$_{3}$SH \citep{maj16, dro18} towards the IRAS 16293 B, we also search the rotational emission and absorption lines of different SH-bearing molecules (see Table~\ref{tab:other molecule}). We cannot detect those molecules using the LTE modelling. We create the integrated emission maps of t-HC(O)SH towards the IRAS 16293 B and we observed the detected emission lines of t-HC(O)SH arise from the warm-inner region of the hot corino IRAS 16293 B. We also discuss the possible formation mechanism of HC(O)SH in the ISM. We discuss that the HC(O)SH molecule may be created either by the reaction between HCO and SH or by the hydrogenation of OCS. We believe that a three-phase warm-up chemical model is needed to understand which formation mechanism is efficient towards the IRAS 16293 B. The successful detection of t-HC(O)SH indicates that a spectral survey is needed using the ALMA for other S and SH-bearing molecules towards the IRAS 16293 B to understand the thiol chemistry in the ISM.

\section*{Acknowledgement}
We thank the anonymous referee for the helpful comments that improved the manuscript. A.M. acknowledges the Swami Vivekananda Merit-cum-Means Scholarship (SVMCM), Government of West Bengal, India, for financial support for this research.\\ This paper makes use of the following ALMA data: ADS /JAO.ALMA\#2016.1.01150.S. ALMA is a partnership of ESO (representing its member states), NSF (USA), and NINS (Japan), together with NRC (Canada), MOST and ASIAA (Taiwan), and KASI (Republic of Korea), in cooperation with the Republic of Chile. The Joint ALMA Observatory is operated by ESO, NRAO, and NAOJ.

\bibliographystyle{aasjournal}
%\bibliography{./literature.bib,added.bib} % if your bibtex file is called example.bib

\end{document}